# Comparative study of spectral broadening and few-cycle compression of Yb:KGW laser pulses in gas-filled hollow-core fibers

**ISLAM SHALABY[1], MICHAEL MCDONNELL[1], COLIN MURPHY[1], NISNAT CHAKRABORTY[1], KODY GRAY[2], JAMES WOOD[2], DIPAYAN BISWAS[1], AND ARVINDER SANDHU[1,2*]**

[1] *Department of Physics, University of Arizona, Tucson, Arizona, 85721, USA*
[2] *College of Optical Sciences, University of Arizona, Tucson, Arizona 85721, USA*
*\*asandhu@arizona.edu*

**Abstract:** While industrial-grade Yb-based amplifiers have become very prevalent, their limited gain bandwidth has created a large demand for robust spectral broadening techniques that allow for few-cycle pulse compression. In this work, we perform a comparative study between several atomic and molecular gases as media for spectral broadening in a hollow-core fiber geometry. Exploiting nonlinearities such as self-phase modulation, self-steepening, and stimulated Raman scattering, we explore the extent of spectral broadening and its dependence on gas pressure, the critical power for self-focusing, and the optimal regime for few-cycle pulse compression. Using a 3-mJ, 200-fs input laser pulses, we achieve ∼ 15 fs, few-cycle pulses with >70% overall energy transmission efficiency. The optimal parameters can be scaled for higher or lower input pulse energies with appropriate gas parameters and fiber geometry.

## 1. Introduction

Modern ultrafast experiments require high-energy, few-cycle pulses to drive high harmonic generation [1–4] and to study electron and nuclear dynamics in atomic, molecular, and solid-state samples [5–8] using time-resolved photon [9] and electron/ion [10] spectroscopies. Conventionally, Ti:sapphire lasers have served as the primary source of short, broadband pulses. More recently, Ytterbium (Yb)-based lasers have introduced a viable alternative to Ti:sapphire lasers; offering higher repetition rates and industrial reliability [11]. The relatively narrow-gain bandwidth of Yb-based lasers, however, limits pulse durations to hundreds of femtoseconds; elevating the demand for bandwidth broadening and pulse compression techniques. Various techniques have been implemented [12] for bandwidth broadening by exploiting the nonlinear response in a medium to generate new frequencies via self-phase modulation or stimulated Raman scattering. These techniques can be used in different geometries such as hollow-core fibers (HCFs) [13–17] or Herriott cells [18, 19], in gaseous media, multi-plate compression (MPC) [20, 21], in solid media, or filamentation [22, 23] in gaseous or solid media.

While each geometry has its own merits, HCFs offer excellent beam profile, due to laser coupling into fiber modes, uniform broadening across beam profile, minimizing spatial chirp, and improved pointing stability [24]. Moreover, stretched HCFs offers additional tunability of fiber lengths and enhanced laser coupling through flexible alignment; ensuring straightness [25]. HCFs' functionality can also be extended to the long wavelength mid-infrared (MIR) regime [26] and can be scalable to high energy lasers [27]. Molecular-gas-filled HCFs can be used to redshift the optical spectrum, translating the central near-infrared (NIR) wavelength to the MIR regime; while achieving broadband, compressible pulses [28]. HCFs can also be used to achieve a soliton state, which can be utilized for temporal self-compression [29], or efficient generation of deep ultraviolet (DUV) pulses through resonant dispersive-wave emission [30, 31]. Short DUV pulses can be utilized to excite molecular transitions, in the 4-6 eV range, in organic samples, for time-resolved measurements. On the other hand, short MIR pulses are also of great importance in the HHG process, in order to extent the harmonic cutoff and generate coherent X-ray pulses [32].

In this work, we investigate and compare the nonlinear response of all atomic gases as well as SF6; commonly used for spectral broadening in a HCF geometry. We conducted experiments and simulations to examine the extent of spectral broadening in each gas species as a function of pressure as well as the temporal compressibility in each case; by providing fixed GDD compensation using chirped mirrors. The critical power for self-focusing is also investigated, for each gas, which imposes limitations on the usability of certain gases for different laser peak powers. Finally, we show optimal gas pressures, for our geometry, to achieve few-cycle (∼ 15 fs) post compression for different gas species.

## 2. Experimental setup

In this work, we devised a HCF solution to spectrally broaden a Yb-based, industrial-grade, 3-mJ laser (Pharos, Light Conversion), then temporally compressed it down to a few-cycle pulse using chirped mirrors. In typical HCF geometries, the fiber input and output ends are mounted on translation stages in order to adjust the coupling of the input laser and optimize the mode and transmission efficiency of the output beam. In contrast to this arrangement, our fiber setup exploits a fixed geometry where the fiber is fixed to the optical table and the input beam is optimally coupled to the fiber input by aligning the beam using a pair of mirrors. The fixed geometry eliminates the substantial cost of the translation stages and improves the mechanical stability resulting in lower pointing and spectral fluctuations.

The 15W Yb:KGW (Pharos, Light Conversion) laser produces a pulse with 200 fs FWHM duration, 3 mJ pulse energy, and a max repetition rate of 5 kHz. The Pharos laser beam diameter $D_{1/e^2}^{in} = 8mm$ and is focused into the fiber using a 2 m lens. With $M^2 \approx 1.2$, the focused beam waist (diameter) is $2\omega_0 \approx 390\mu m$. We employ a fused silica based stretched hollow-core fiber (Polymicro-TSP530700) with an inner (hollow core) diameter ($D_{core} = 530\mu m$) and an outer (cladding) diameter ($D_{cladding} = 700\mu m$). To ensure straightness, the fiber is first stretched through the mounting apertures using stretching posts. The stretching posts positioning can be adjusted to ensure a straight path between the apertures. An initial alignment is performed using a HeNe laser to ensure straightness prior to fixing the fiber. Once the fiber is aligned and stretched, it can be glued to the apertures on both ends to provide a vacuum seal. The fiber is later scored on both ends using a Ruby fiber scribe (THORLABS-S90R) to ensure a clean cut for optimal laser coupling. The apertures are finally connected to KF tubes on both ends to form an enclosure for the filled gas. The main laser can be aligned beforehand in air for easier adjustment. Once the transmission and mode of the fiber output are optimized, the enclosures can be sealed and filled with various research-grade gases at different pressures. The fiber output is directed towards chirped mirrors (Ultrafast Innovations-PC1611) providing a nominal GDD of $-150 fs^2$ per bounce. The beam experiences 5 bounces in total; providing a total GDD value of $-750 fs^2$. After chirped mirrors, the beam passes through variable-thickness fused-silica wedges for GDD fine tuning; providing a GVD value of $+57 fs^2/mm$.

The output beam is recollimated using a 1.5m lens, which telescopes the beam diameter down to $D_{1/e^2}^{out} = (f_{out}/f_{in}) \cdot D_{1/e^2}^{in} = 6mm$. The smaller output beam diameter allows for multiple bounces per chirped mirror, reducing the number of used chirped mirrors. The spectrum is analyzed using a homemade IR spectrometer with a 300 g/mm grating which is calibrated using Neon lamp IR emission lines. The pulse duration is characterized using a homemade SHG-FROG apparatus with all-reflection geometry which eliminates any added dispersion from beamsplitters in a typical SHG-FROG arrangement. In our FROG apparatus, a mask clips the input beam creating two adjacent pencil beams which reflect off from two concave, D-shaped mirrors with $f = 10cm$. The focused beams are matched spatiotemporally on a thin ($10\mu m$) BBO crystal to produce a second harmonic signal. The signal beam is spatially filtered using a second mask, collimated using a lens, and send to a visible spectrometer (Photon Control, SPM-002). The second IR pencil beam is delay-controlled by linearly motorizing one concave mirror and the resulting spectrogram (FROG trace) is recorded.

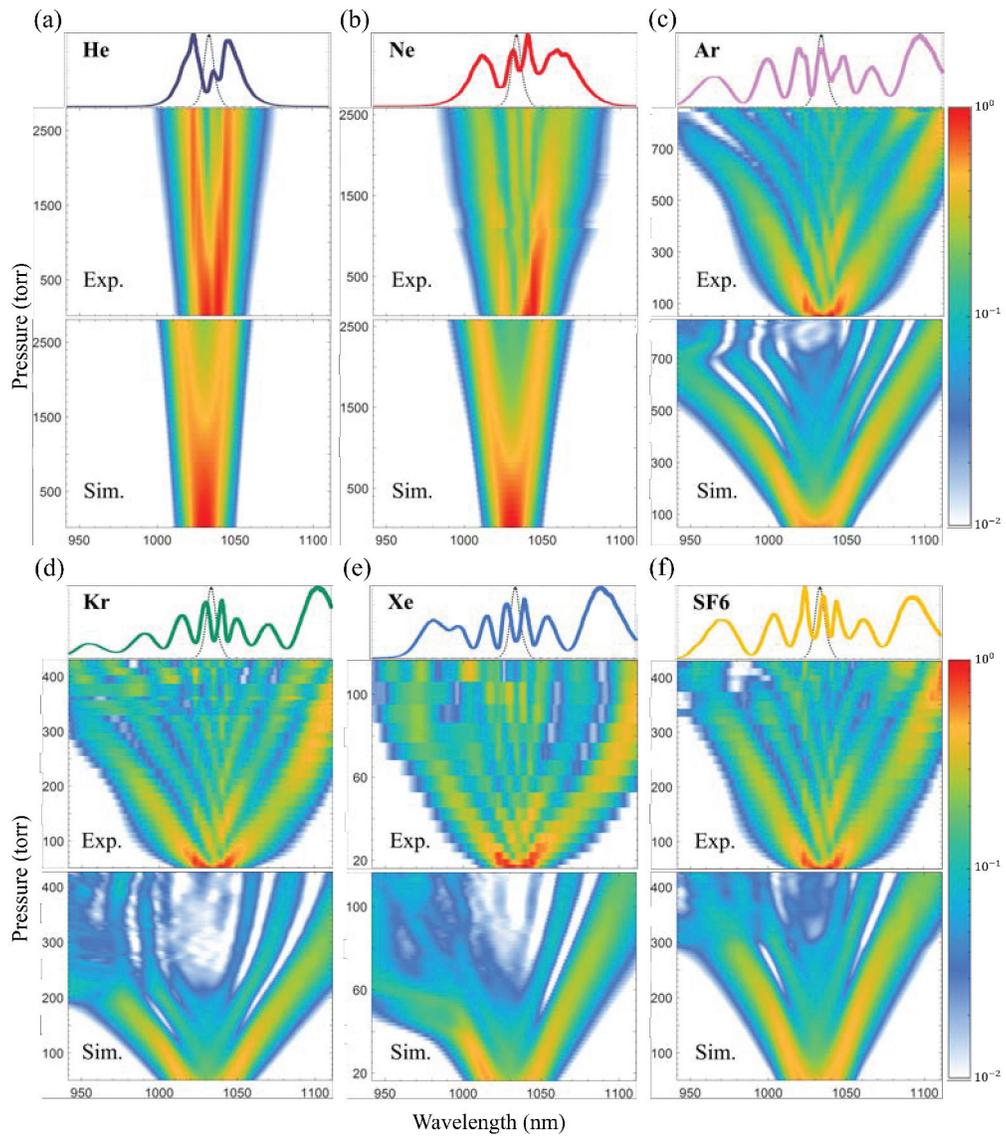

Fig. 1. Middle and bottom panels respectively show experimental and simulated pressure dependence on spectral broadening in **(a)** helium, **(b)** neon, **(c)** argon, **(d)** krypton, **(e)** xenon, and **(f)** SF$_6$. Top panel shows spectrum at pressures resulting in shortest achieved pulses listed in table 1 relative to input spectrum (black, dashed line).

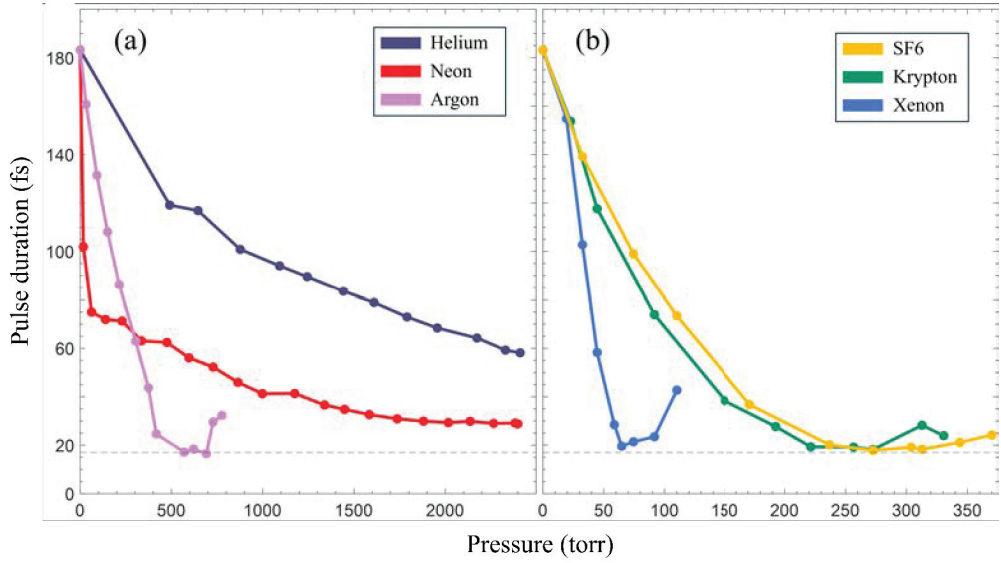

Fig. 2. FWHM pulse duration as a function of gas pressure for **(a)** helium, neon, and argon and **(b)** krypton, xenon, and SF6. Dashed horizontal line at 17 fs represents lowest achieved pulse duration. Due to pressure controller limitations, the maximum gas pressure we were able to produce was approximately 2800 torr which was insufficient pressure for neon and helium to reach 17 fs pulse.

## 3. Theoretical Methods

We used the open-source code Luna [33]. It can simulate a variety of nonlinear optical processes, but it is optimized for ultrafast pulse propagation in fibers. It accounts for linear and fiber modal dispersion, self-phase modulation, self-steepening, self-focusing, photo-ionization, plasma effects, as well as molecular responses, such as stimulated Raman scattering (SRS), resulting from vibrational and rotational couplings. The code projects the frequency-domain electric field, calculated from input beam parameters, onto the transverse modes of the fibers. Each mode is propagated given the medium dispersion, losses, and nonlinear polarization effects, in each propagation step. The final field can be calculated by appropriately summing over the final fiber modes, the spectral intensity and phase can be calculated and Fourier-transformed to give the final temporal field intensity and phase.

The Pharos laser output has a central wavelength of 1030 nm with bandwidth $\Delta\lambda_{FWHM} = 7.7 nm$. The focused high intensity (fluence $F_0 = 5\ J/cm^2$, peak intensity $I_0 = 23\ TW/cm^2$) of the laser field is maintained across the length of the fiber (2.6 m) and it drives spectral broadening due to self-phase modulation (SPM) in the gaseous media. The instantaneous frequency of the laser pulse during propagation can be iteratively calculated using:

$$\omega(t) = \omega_0 \left(1 - \frac{n_2 z}{c} \frac{\partial I(t)}{\partial t}\right) \qquad (1)$$

where $z$ is the propagation step, $\omega_0$ is the central frequency, $c$ is the speed of light, and $n_2$ is the non-linear refractive index ($n(t) = n_o + n_2 I(t)$).

The non-linear refractive index, $n_2$, depends on the polarization response of the medium. For gases, the gas species and pressure dictates the strength of the non-linear response, which is related to the third-order nonlinear susceptibility, $\chi^{(3)}$, [34] by:

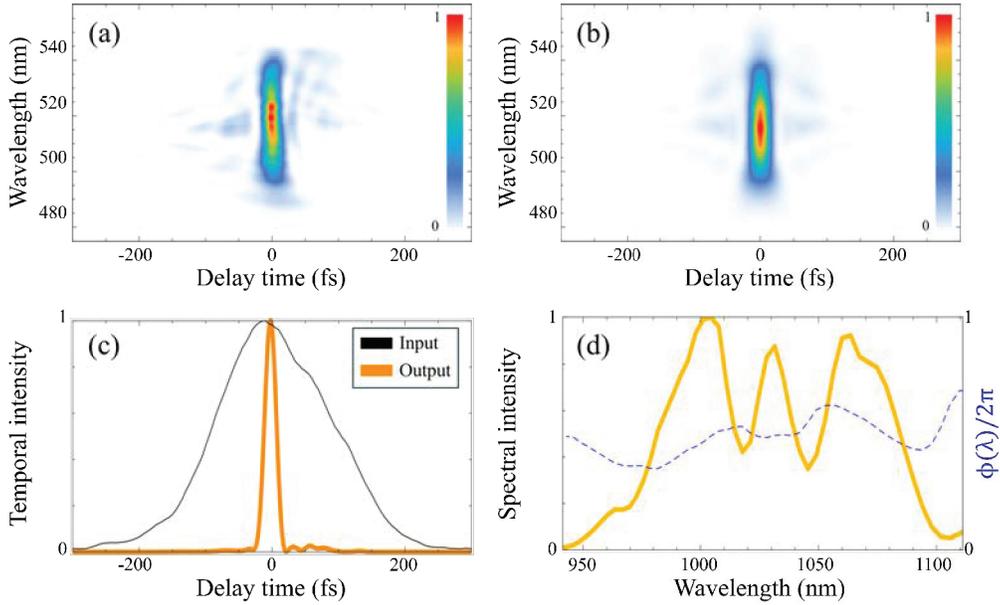

Fig. 3. **(a)** Experimental SHG Frog trace for SF6-filled HCF at optimal pressure $P^*$ (table 1). **(b)** Simulated SHG Frog trace. **(c)** Retrieved temporal profile of uncompressed, input pulse (black line) and compressed, output pulse (dark orange line). The compressed pulse has a FWHM duration of 17 fs. **(d)** Retrieved spectrum (light orange line) and spectral phase (blue, dashed line).

$$n_2 = \frac{3}{4} \frac{\chi^{(3)}}{\epsilon_0 c n_0^2} \tag{2}$$

Historically, The nonlinear susceptibility $\chi^{(3)}$, has been measured by means of nonlinear harmonic generation [35, 36]. Since the nonlinear response in gases is density-dependent, the measured values for third-order susceptibility $\chi^{(3)}$ are measured at STP ($P_0$ =1 atm, $T_0 = 273K$ [0° C]), and can be scaled, for any temperature and pressure, approximately, as follow:

$$\chi^{(3)}(P,T) \approx \frac{\tilde{N}(P,T)}{\tilde{N}(P_0,T_0)} * \chi^{(3)}(P_0,T_0) = \frac{P}{P_0}\frac{T_0}{T} * \chi^{(3)}(P_0,T_0) \tag{3}$$

where $\tilde{N} = P/K_B T$ is the number density. The nonlinear response can also depend on the driving field wavelength $\lambda_0$. Most measurements have been performed for $\lambda_0 \approx 800nm$ and $\lambda_0 \approx 1064nm$ [35, 36], and, in this work, the slight variation resulting from wavelength dependence has been ignored since our central wavelength ($\lambda_0 = 1030nm$) is close to the central wavelength in the reported measurements.

The nonlinear refractive index $n_2$ also depends on the linear refractive index $n_0$ (eq. [2]); which also depends on pressure and temperature. This dependence can be modeled in different ways, and for this work we adopt a transformation based on Sellmeier coefficients [37], as follows:

$$n_0^2(\lambda, P, T) - 1 = \frac{P}{P_0}\frac{T_0}{T} * \left[ \frac{B_1 \lambda^2}{\lambda^2 - C_1} + \frac{B_2 \lambda^2}{\lambda^2 - C_2} \right]_{P_0, T_0} \tag{4}$$

Finally, the spatial Kerr effect (self-focusing) has to also be taken into account. Surpassing the critical power for self-focusing in the fiber can lead to ionization and beam defocusing, which disrupts the SPM process resulting in distortions in the spectral broadening. The critical power for self-focusing $P_{crit.}$ can be estimated [38] as follows:

$$P_{crit.} = \frac{0.148\lambda^2}{n_0 n_2} \qquad (5)$$

For our laser system, The peak power $P_{peak} = E_{pulse}/\tau_{pulse}$ = 3 mJ / 200 fs = 15 GW. To ensure that the peak power is lower that the critical power for each gas species, $n_2 < 1e-23\ m^2$/W ($\lambda$ = 1030 nm) for $P_{crit.} \approx 15.7$ GW.

## 4. Results and discussion

Figure 1 shows the spectral broadening as a function of gas pressure for different gas species. For each gas, we have a separate panel showing optimal spectrum (top), experimental variation of spectrum with pressure (middle) and simulation results (bottom). In general, the experimental results and simulations agree very well for the entire range of experimentation. For He and Ne, the spectral broadening is minimal due to their low nonlinear response. Even though these gas species are not practical for our pulse energy (3 mJ), they are convenient for high pulse-energy laser systems (> 5 mJ) or smaller fiber geometries (Dia. < 250 $\mu$m); especially their high resistance to self-focusing can accommodate high intensities at high pressures. For intermediate pulse energies (1-3 mJ), Ar, Kr, and SF6 are ideal to produce large spectral broadening at moderate pressures as shown in figure 1(c),(d),(f). It is important to choose convenient gas species and fiber geometry to achieve large bandwidths around moderate pressures (~ 1 atm) to ensure mechanical stability of the fiber; which affects spectral and mode stability.

As per our calculations, the critical limit for self-focusing occurs at pressures of 21,000 for He, 11,700 for Ne, 900 torr for Ar, 330 torr for Kr, 110 torr for Xe, and 960 torr for SF6. These values agree with experimental and simulation results (figure 1(c),(d),(e)) where the smooth spectral broadening starts to distort around these pressure for each gas species. Calculated pressure for critical power in SF6 does not match the simulation and experimental results in figure 1(f) as only Kerr effect from electronic response is considered for this calculation of critical power. Molecular couplings can contribute to the nonlinear effects which is observed experimentally and is taken into account in the simulations, as discussed later.

Figure 1(d),(e), show the rapid spectral broadening in Kr and Xe, respectively, as a function of pressure. Due to the variation of $n_2$ with pressure, and hence the reduction in the critical power for self-focussing, the spectral distortion sets in at pressures of ~ 330 torrs for Kr, and 110 torrs for Xe. Therefore, despite achieving short pulses (~ 17 fs) in Kr and Xe for our system, as shown in table 1, the optimal pressures are much lower than atmospheric pressure; which has a potential drawback that over time atmosphere can diffuse in and the reduction in gas purity can disrupt the SPM process, resulting in spectral and pulse duration instabilities. These gas species are convenient for low pulse-energy laser systems (< 1 mJ) and can be used to produce large bandwidths; enough to achieve few-cycle pulses [15].

Figure 1.(f) shows the broadening in SF6-filled HCF which resembles the amount of broadening in Kr, despite having lower nonlinear index, $n_2$, for the same pressure ($P^*$) (table 1). The calculated nonlinear indices are purely derived from electronic responses; which is most accurate for atomic gases. Molecular gases have additional vibrational and rotational couplings which can enhance spectral broadening through stimulated Raman scattering (SRS) [11, 28, 39, 40]. In the SF6 case, the rotational response is minimal due to the symmetry of the SF6 molecule and only vibrational modes are considered in the simulations. While the electronic response for SF6 is comparable to Argon [39], additional molecular couplings effectively improves spectral broadening where the overall response is comparable to Krypton.

Figure 2 shows pulse durations for different gas species as a function of pressure; after chirped-mirror compression and measured using SHG-FROG. For He and Ne in figure 2(a), optimal pulse durations could not be achieved due to high-pressure limitations of our equipment. For Xe, the output pulse duration is highly sensitive to changes in the gas pressure. This results from the high dependence of the output spectrum on small pressure variations (figure 1(e)). Figure 2(b) shows the similarity between Kr and SF6 in the final pulse durations due to similar broadening effects as discussed earlier.

| Gas | $\chi^{(3)}/\chi^{(3)}_{(He)}$ | $P^*$ (torr) | $\tau(P^*)$ (fs) | $n_2(P^*)(\frac{pm^2}{W})$ | $n_0(P^*)$ |
|---|---|---|---|---|---|
| Helium | 1 | - | - | - | - |
| Neon | 1.8 | - | - | - | - |
| Argon | 23.5 | 690 | 16.6 | 7.71 | 1.000236 |
| Krypton | 64 | 275 | 18.2 | 8.29 | 1.000141 |
| Xenon | 188.2 | 65 | 19.6 | 5.78 | 1.000054 |
| SF6 | 32.3 | 275 | 17.3 | 3.11[†] | 1.162049 |

Table 1. Relative nonlinear susceptibility [35] compared to He, $\chi^{(3)}_{(He)} = 0.00126$ pm$^2$/V$^2$ (at 1 atm, 25°C). This value is a factor of 4 larger than [35] because of their definition of hyperpolarizability. Pressures $P^*$ are values where shortest pulse durations $\tau(P^*)$ are found for different gas species. $n_0(P^*)$ and $n_2(P^*)$ are the respective linear and nonlinear indices at optimal pressures $P^*$. [†] $n_2$ value for SF6 is calculated using only Kerr effect from electronic response and does not account for molecular contributions.

Considering that SF6 can achieve a short pulse for a wider range of pressures (figure 2(b)) and is also more cost-effective, it makes it a better candidate for pulse compression applications. For Ar (figure 2(a)), short pulse durations can be achieved for a good range of pressures, similar to SF6, but at pressures closer to 1 atm, which improves mechanical stability, making Ar the best candidate; given its cost-effectiveness as well. In table 1, relative nonlinearities of different gas species are listed with their linear and nonlinear indices at optimal pressures $P^*$ where shortest pulse durations are observed.

Finally, we show the pulse characterization results in figure 3(a), we show the experimental FROG trace obtained in SF6 at the optimal pressure . The FROG reconstruction is shown in figure 3(b) and the temporal and spectral intensities and phases obtained are plotted in figure 3(c),(d). We observe a clean 17fs pulse generation with almost negligible temporal structure around it. We observe that once properly setup, such an operation can be sustained for weeks without the need for realignment or adjustment.

## 5. Conclusion

In this study, we compared the nonlinear properties of all atomic gases as well as SF6 and their utility in broadband, few-cycle pulse generation. We explored the pressure dependence of spectral broadening, critical power, and pulse compression. These results can be mapped to different laser intensities, fiber geometries, and can even guide other spectral broadening methods. We Show the generation of broadband spectra in different gas species and the ability of post-compression down to few cycle pulses. While these results are produced with input pulses in the intermediate pulse energy (3 mJ) regime, the same technique can be applied to low pulse-energy (< 1 mJ) lasers [15, 16] and high pulse-energy lasers (>5 mJ) [27, 28]; by selecting a convenient gas species and appropriately scaling fiber geometry [41] to accommodate the input laser intensity.

**Funding.** National Science Foundation award number PHY 2207641.

**Acknowledgments.** This work was supported by the National Science Foundation award number PHY 2207641 and U.S. Department of Energy, Office of Science, Basic Energy Sciences under Award No. DE-SC0018251. We thank Prof. Thomas Weinacht and Prof. Miroslav Kolesik for discussions.

**Disclosures.** The authors declare no conflicts of interest.

**Data availability.** Data underlying the results presented in this paper are not publicly available at this time but may be obtained from the authors upon reasonable request.
## References

1. F. Krausz and M. Ivanov, "Attosecond physics," Rev. Mod. Phys. **81**, 163–234 (2009).
2. P. Antoine, A. L'Huillier, and M. Lewenstein, "Attosecond pulse trains using high–order harmonics," Phys. Rev. Lett. **77**, 1234–1237 (1996).
3. P. B. Corkum, "Plasma perspective on strong field multiphoton ionization," Phys. Rev. Lett. **71**, 1994–1997 (1993).
4. J. L. Krause, K. J. Schafer, and K. C. Kulander, "High-order harmonic generation from atoms and ions in the high intensity regime," Phys. Rev. Lett. **68**, 3535–3538 (1992).
5. A. H. Zewail, "Femtochemistry: Atomic-scale dynamics of the chemical bond," The J. Phys. Chem. A **104**, 5660–5694 (2000).
6. E. Goulielmakis, Z.-H. Loh, A. Wirth, R. Santra, N. Rohringer, V. S. Yakovlev, S. Zherebtsov, T. Pfeifer, A. M. Azzeer, M. F. Kling, S. R. Leone, and F. Krausz, "Real-time observation of valence electron motion," Nature **466**, 739–743 (2010).
7. M. Uiberacker, T. Uphues, M. Schultze, A. J. Verhoef, V. Yakovlev, M. F. Kling, J. Rauschenberger, N. M. Kabachnik, H. Schröder, M. Lezius, K. L. Kompa, H.-G. Muller, M. J. J. Vrakking, S. Hendel, U. Kleineberg, U. Heinzmann, M. Drescher, and F. Krausz, "Attosecond real-time observation of electron tunnelling in atoms," Nature **446**, 627–632 (2007).
8. M. Schultze, K. Ramasesha, C. Pemmaraju, S. Sato, D. Whitmore, A. Gandman, J. S. Prell, L. J. Borja, D. Prendergast, K. Yabana, D. M. Neumark, and S. R. Leone, "Attosecond band-gap dynamics in silicon," Science **346**, 1348–1352 (2014).
9. I. Shalaby, N. Chakraborty, S. Yanez-Pagans, J. Wood, D. Biswas, and A. Sandhu, "Probing ultrafast excited-state dynamics using EUV-IR six-wave-mixing emission spectroscopy," Opt. Express **30**, 46520–46527 (2022).
10. D. Biswas, J. K. Wood, I. Shalaby, and A. Sandhu, "Ultrafast dynamics of the rydberg states of $CO_2$: Autoionization and dissociation lifetimes," Phys. Rev. A **110**, 043106 (2024).
11. T.-C. Truong, J. E. Beetar, and M. Chini, "Light-field synthesizer based on multidimensional solitary states in hollow-core fibers," Opt. Lett. **48**, 2397 (2023).
12. T. Nagy, P. Simon, and L. Veisz, "High-energy few-cycle pulses: post-compression techniques," Adv. Physics: X **6**, 1845795 (2021).
13. M. Nisoli, S. De Silvestri, and O. Svelto, "Generation of high energy 10 fs pulses by a new pulse compression technique," Appl. Phys. Lett. **68**, 2793–2795 (1996).
14. M. Nisoli, "Hollow fiber compression technique:a historical perspective," IEEE J. Sel. Top. Quantum Electron. p. 1–14 (2024).
15. J. E. Beetar, F. Rivas, S. Gholam-Mirzaei, Y. Liu, and M. Chini, "Hollow-core fiber compression of a commercial Yb:KGW laser amplifier," J. Opt. Soc. Am. B **36**, A33 (2019).
16. Z. Pi, H. Y. Kim, and E. Goulielmakis, "Petahertz-scale spectral broadening and few-cycle compression of Yb:KGW laser pulses in a pressurized, gas-filled hollow-core fiber," Opt. Lett. **47**, 5865 (2022).
17. J. Codere, M. Belmonte, B. Kaufman, M. Wahl, E. Jones, M. G. Cohen, T. Weinacht, and R. Forbes, "High repetition-rate pulse shaping of a spectrally broadened Yb femtosecond laser," Opt. Continuum **3**, 785 (2024).
18. M. Kaumanns, V. Pervak, D. Kormin, V. Leshchenko, A. Kessel, M. Ueffing, Y. Chen, and T. Nubbemeyer, "Multipass spectral broadening of 18 mJ pulses compressible from 13 ps to 41 fs," Opt. Lett. **43**, 5877 (2018).
19. P. Balla, A. Bin Wahid, I. Sytcevich, C. Guo, A.-L. Viotti, L. Silletti, A. Cartella, S. Alisauskas, H. Tavakol, U. Grosse-Wortmann, A. Schönberg, M. Seidel, A. Trabattoni, B. Manschwetus, T. Lang, F. Calegari, A. Couairon, A. L'Huillier, C. L. Arnold, I. Hartl, and C. M. Heyl, "Postcompression of picosecond pulses into the few-cycle regime," Opt. Lett. **45**, 2572 (2020).
20. C.-H. Lu, Y.-J. Tsou, H.-Y. Chen, B.-H. Chen, Y.-C. Cheng, S.-D. Yang, M.-C. Chen, C.-C. Hsu, and A. H. Kung, "Generation of intense supercontinuum in condensed media," Optica **1**, 400 (2014).
21. J. Li, W. Tan, J. Si, Z. Kang, and X. Hou, "Generation of ultrabroad and intense supercontinuum in mixed multiple thin plates," Photonics **8**, 311 (2021).
22. C. Hauri, W. Kornelis, F. Helbing, A. Heinrich, A. Couairon, A. Mysyrowicz, J. Biegert, and U. Keller, "Generation of intense, carrier-envelope phase-locked few-cycle laser pulses through filamentation," **79**, 673–677.
23. F. Silva, D. R. Austin, A. Thai, M. Baudisch, M. Hemmer, D. Faccio, A. Couairon, and J. Biegert, "Multi-octave supercontinuum generation from mid-infrared filamentation in a bulk crystal," **3**, 807. Publisher: Nature Publishing Group.
24. L. Gallmann, T. Pfeifer, P. Nagel, M. Abel, D. Neumark, and S. Leone, "Comparison of the filamentation and the hollow-core fiber characteristics for pulse compression into the few-cycle regime," **86**.

25. T. Nagy, M. Forster, and P. Simon, "Flexible hollow fiber for pulse compressors," **47**. Publisher: Optica Publishing Group.
26. G. Fan, T. Balčiūnas, T. Kanai, T. Flöry, G. Andriukaitis, B. E. Schmidt, F. Légaré, and A. Baltuška, "Hollow-core-waveguide compression of multi-millijoule CEP-stable 3.2-$\mu$m pulses," **3**. Publisher: Optica Publishing Group.
27. G. Fan, P. A. Carpeggiani, Z. Tao, G. Coccia, R. Safaei, E. Kaksis, A. Pugzlys, F. Légaré, B. E. Schmidt, and A. Baltuška, "70 mJ nonlinear compression and scaling route for an Yb amplifier using large-core hollow fibers," **46**. Publisher: Optica Publishing Group.
28. P. A. Carpeggiani, G. Coccia, G. Fan, E. Kaksis, A. Pugžlys, A. Baltuška, R. Piccoli, Y.-G. Jeong, A. Rovere, R. Morandotti, L. Razzari, B. E. Schmidt, A. A. Voronin, and A. M. Zheltikov, "Extreme Raman red shift: ultrafast multimode nonlinear space-time dynamics, pulse compression, and broadly tunable frequency conversion," Optica **7**, 1349 (2020).
29. J. C. Travers, "Optical solitons in hollow-core fibres," Opt. Commun. **555**, 130191 (2024).
30. J. C. Travers, T. F. Grigorova, C. Brahms, and F. Belli, "High-energy pulse self-compression and ultraviolet generation through soliton dynamics in hollow capillary fibres," Nat. Photonics **13**, 547–554 (2019). Publisher: Nature Publishing Group.
31. C. Brahms, F. Belli, and J. C. Travers, "Resonant dispersive wave emission in hollow capillary fibers filled with pressure gradients," Opt. Lett. **45**, 4456 (2020).
32. T. Popmintchev, M.-C. Chen, D. Popmintchev, P. Arpin, S. Brown, S. Ališauskas, G. Andriukaitis, T. Balčiunas, O. D. Mücke, A. Pugzlys, A. Baltuška, B. Shim, S. E. Schrauth, A. Gaeta, C. Hernández-García, L. Plaja, A. Becker, A. Jaron-Becker, M. M. Murnane, and H. C. Kapteyn, "Bright coherent ultrahigh harmonics in the keV X-ray regime from Mid-Infrared femtosecond lasers," Science **336**, 1287–1291 (2012).
33. C. Brahms and J. C. Travers, "Luna.jl," https://doi.org/10.5281/zenodo.5797786.
34. R. W. Boyd, *Nonlinear Optics* (Academic, 2008), 3rd ed.
35. H. J. Lehmeier, W. Leupacher, and A. Penzkofer, "Nonresonant third order hyperpolarizability of rare gases and $N_2$ determined by third harmonic generation," Opt. Commun. **56**, 67–72 (1985).
36. D. P. Shelton, "Nonlinear-optical susceptibilities of gases measured at 1064 and 1319 nm," Phys. Rev. A **42**, 2578–2592 (1990).
37. E. T. J. Nibbering, G. Grillon, M. A. Franco, B. S. Prade, and A. Mysyrowicz, "Determination of the inertial contribution to the nonlinear refractive index of air, $N_2$, and $O_2$ by use of unfocused high-intensity femtosecond laser pulses," J. Opt. Soc. Am. B **14**, 650 (1997).
38. G. Fibich and A. L. Gaeta, "Critical power for self-focusing in bulk media and in hollow waveguides," Opt. Lett. **25**, 335–337 (2000).
39. A. Hoffmann, M. Zürch, and C. Spielmann, "Extremely nonlinear optics using shaped pulses spectrally broadened in an Argon- or Sulfur hexafluoride-filled hollow-core fiber," Appl. Sci. **5**, 1310–1322 (2015).
40. S.-F. Gao, Y.-Y. Wang, F. Belli, C. Brahms, P. Wang, and J. C. Travers, "From Raman frequency combs to supercontinuum generation in nitrogen-filled hollow-core anti-resonant fiber," Laser & Photonics Rev. **16**, 2100426 (2022).
41. C. M. Heyl, H. Coudert-Alteirac, M. Miranda, M. Louisy, K. Kovacs, V. Tosa, E. Balogh, K. Varjú, A. L'Huillier, A. Couairon, and C. L. Arnold, "Scale-invariant nonlinear optics in gases," Optica **3**, 75–81 (2016).